# Time, Topology and the Twin Paradox

Jean-Pierre Luminet

*Laboratoire Univers et Théories, CNRS-UMR 8102,*
*Observatoire de Paris, F-92195 Meudon cedex, France*
jean-pierre.luminet@obspm.fr

**Summary**

The twin paradox is the best known thought experiment associated with Einstein's theory of relativity. An astronaut who makes a journey into space in a high-speed rocket will return home to find he has aged less than a twin who stayed on Earth. This result appears puzzling, since the situation seems symmetrical, as the homebody twin can be considered to have done the travelling with respect to the traveller. Hence it is called a "paradox". In fact, there is no contradiction and the apparent paradox has a simple resolution in Special Relativity with infinite flat space. In General Relativity (dealing with gravitational fields and curved space-time), or in a compact space such as the hypersphere or a multiply connected finite space, the paradox is more complicated, but its resolution provides new insights about the structure of spacetime and the limitations of the equivalence between inertial reference frames.

**Play time**

The principle of relativity ensures equivalence between inertial reference frames in which the equations of mechanics "hold good" in their simplest form. Inertial frames are spatial coordinate systems together with some means of measuring time so that observers attached to them can distinguish uniform motions from accelerated motions.

In classical mechanics as well as in Special Relativity, such privileged frames are those moving at a constant velocity, i.e. in uniform rectilinear motion. Their rest states are equivalent, as every passenger in a train slowly starting relative to a neighboring one at a train station can check. Without feeling any acceleration, the passenger cannot decide which train is moving with respect to the other one.

Classical mechanics makes the assumption that time flows at the same rate in all inertial reference frames. As a consequence, the mathematical transformations between inertial systems are just the usual Galilean formulae, which preserve space intervals $\Delta d$ and time intervals $\Delta t$. As invariant quantities, lengths and durations are independent of the positions and speeds of the reference frames in which they are measured. This corresponds to Newton's concepts of absolute space and absolute time.

Special Relativity makes a different assumption, namely that the speed of light in vacuum, $c$, remains the same for every observer, whatever his state of motion. This assumption was confirmed by the famous Michelson and Morley experiments (1887). The mathematical transformations between inertial systems are given by the Lorentz formulae, which allow us to reformulate the laws of mechanics and electromagnetism in a coherent way. Their most immediate consequence is that space and time are not absolute but "elastic", in the sense that space intervals $\Delta d$ and time intervals $\Delta t$ now depend on the relative velocity between the observer and the system he measures.

However, the Lorentz transformations preserve the space-time interval, an algebraic combination of space and time intervals given by $\Delta s = \sqrt{c^2 \Delta t^2 - \Delta d^2}$. According to the Lorentz formulae, the clock of a system in motion appears to tick slower than that of a system at rest, while distances in the moving system appear to be shortened. In effect, for an observer at rest with his clock $\Delta d = 0$, and Δ*s* measures his so-called *proper time*. But if an observer moves relative to a clock, he will measure a time interval Δ*t* longer and a space interval Δ*d* shorter than the observer at rest. These rather counter-intuitive effects are called *apparent time dilation* (moving clocks tick more slowly) and *length contraction* (moving objects appear shortened in the direction of motion).

The more the relative velocity *v* increases, the more the clock appears to slow down. Due to the expression of the coefficient of time dilation, $\sqrt{(1-v^2/c^2)}$, the phenomenon is noticeable only at velocities approaching that of light. At the extreme limit, for a clock carried by a photon, which is a particle of light, time does not flow at all. The photon never ages, because its proper time Δ*s* is always zero.

Special Relativity is one of the best verified theories in physics. The reality of apparent time dilation has been tested experimentally using elementary particles that can be accelerated to velocities very close to the speed of light. For instance, muons, unstable particles which disintegrate after 1.5 microseconds of proper time, were accelerated until they reached 0.9994 *c*. Their apparent lifetime (as measured in the rest frame of the laboratory) extended to 44 microseconds, which is thirty times longer than their real lifetime, in complete agreement with special relativistic calculations.

In order to avoid misunderstanding, it is very important to make a distinction between the apparent time and the proper time. To illustrate the difference, let us compare two identical clocks consisting of light impulses traveling between two parallel mirrors. One of the clocks is in uniform rectilinear motion at velocity *v* relative to the other one (in a direction parallel to the line joining the mirrors). At moment *t* = 0, both clocks are at the same location, and the light impulses are sent to each of them. At time *t*, the observer of the clock at rest checks that the beam of light reaches the second mirror – this moment corresponds to the first tick of the clock. The second clock moved during this time, and the beam of light has yet to reach its second mirror. Thus it seems to run more slowly, because its ticks are not synchronized with those of the clock at rest. But, as the notion of uniform motion is a relative one, *these effects are totally symmetric*. The observer bound to the clock in motion considers himself at rest, and he sees the other system moving. He thus sees the clock of the other system slowing down.

In other words, the observers perform observations which apparently contradict one another, each seeing the other clock beating more slowly than his own. However, their points of view are not incompatible, because this *apparent* dilation of time is an effect bound to observation, and the Lorentz transformation formulae ensure the coherence of both measurements. Indeed, in the case of uniform rectilinear relative motion, the *proper* times of both clocks remain perfectly identical; they "age" at the same rate.

**The twin paradox in Special Relativity**

Now consider two clocks brought together in the same inertial reference frame and synchronised. What happens if one clock moves away in a spaceship and then returns? In his seminal paper on Special Relativity, Albert Einstein (see Einstein 1905) predicted that the clock that undergoes the journey would be found to lag behind the clock which stays put. Here the time delay involves the proper time, not the apparent one. To emphasize on this, in 1911 Einstein restated and elaborated on this result in the following statement: "If we placed a living organism in a box... one could arrange that the organism, after any arbitrary lengthy flight, could be returned to its original spot in a scarcely altered condition, while corresponding organisms which had remained in their original positions had already long since given way to new generations. For the moving organism the lengthy time of the journey was a mere instant, provided the motion took place with approximately the speed of light." (in Resnick and Halliday 1992)

The same year, the French physicist Paul Langevin (see Langevin 1911) picturesquely formulated the problem using the example of twins aging differently according to their respective space-time trajectories (called *worldlines*). One twin remains on Earth while the other undertakes a long space journey to a distant planet, in a spaceship moving at almost the speed of light, then turns around and returns home to Earth. There the astronaut discovers that he is younger than his sibling. That is to say, if the twins had been carrying the clocks mentioned above, the traveller's clock would be found to lag behind the clock which stayed with the homebody brother, meaning that less time has elapsed for the traveller than for the homebody. This result indeed involves *proper times* as experienced by each twin, since biological clocks are affected in the same way as atomic clocks. The twins' ages can also be measured in terms of the number of their heartbeats. The traveller is really younger than his homebody twin when he returns home.

However, due to apparent time dilation, each twin believes the other's clock runs slower, and so the paradox arises that each believes the other should be younger at their reunion. In other words, the symmetry of their points of view is broken. Is this paradoxical?

In scientific usage, a paradox refers to results which are contradictory, i.e. logically impossible. But the twin paradox is not a logical contradiction, and neither Einstein nor Langevin considered such a result to be literally paradoxical. Einstein only called it "peculiar", while Langevin explained the different aging rates as follows: "Only the traveller has undergone an acceleration that changed the direction of his velocity." He showed that, of all the worldlines joining two events (in this example the spaceship's departure and return to Earth), the one that is not accelerated takes the longest proper time.

The twin paradox, also called the Langevin effect, underlines a limitation of the principle of relativity: points of view are symmetrical only for inertial reference systems, i.e. those that aren't undergoing any acceleration. In the twin experiment, the Earth and the spaceship are not in a symmetrical relationship; the ship has a "turnaround" in which it undergoes non-inertial motion, while the Earth has no such turnaround. Since there is no symmetry, Special Relativity is not contradicted by the realization that the twin who left Earth is younger than his sibling at the time of their reunion. The subject has been widely discussed for pedagogical purposes, see e.g. Taylor and Wheeler (1992)

**An example with numbers**

Let us call the twins Homebody and Traveller. At time $t = 0$ they synchronise their clocks in the Earth's inertial reference frame. Then Homebody stays on Earth whereas Traveller leaves towards a star E situated 10 light years away, travelling at $v = 0.9$ c, that is 270 000 km/s. Next, he returns to Earth with speed $-v$. For convenience the ship is assumed to have instaneous accelerations, so it immediately attains its full speed upon departure, turn around and arrival.

What would each twin observe about the other during the trip? The (x-t) space-time diagrams below (figs. 1-5) allow us to solve the problem without any numerical calculation. We can choose the light-year as the unit of distance and the year as the unit of time. Then the paths of light rays are lines tilted at 45° (the dotted lines). They carry the images of each twin and his age-clock to the other twin. The vertical thick line is Homebody's path through space-time, and Traveller's trajectory (thin line) is necessarily tilted by less than 45° with respect to the vertical. Each twin transmits light signals at equal intervals according to his own clock, but according to the clock of the twin receiving the signals they are not being received at equal intervals.

In this example the coefficient of time dilation is $\sqrt{(1-v^2/c^2)} = 0.436$, that is, when Homebody reads "1 second" on his clock, he reads "0.436 second" on Traveller's clock which is moving away from him at $0.9\ c$, and vice versa.

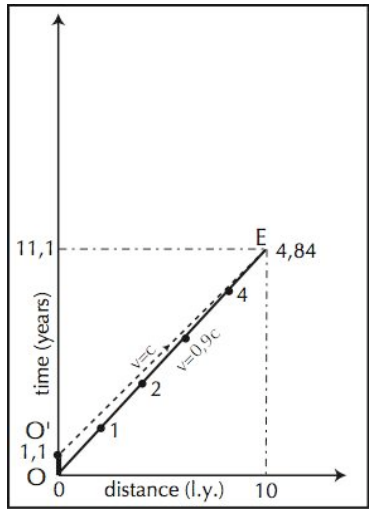

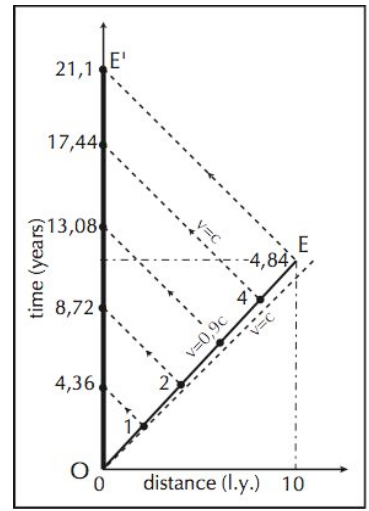

**Fig. 1 Outward Journey: What Traveller measures.** In principle, 11.1 years are required to cover 10 light years at the speed of $0.9\ c$. However, according to his clock, Traveller reaches E after only 4.84 years (11.1 x 0.436). Besides, once he arrives at E Traveller sees the Earth such as it was at O', which is 1.1 years after the departure according to Homebody's clock.
**Conclusion: Traveller sees Homebody's clock beating 4.36 times more slowly.**

**Fig. 2 Outward Journey : What Homebody measures.** Homebody knows that, after 11.1 years, Traveller should arrive at E. However, the light rays sent from E take 10 years to reach him at E'. Homebody thus sees Traveller arriving at E only after 21.1 years.
**Conclusion: Homebody sees Traveller's clock beating 4.36 times more slowly.**

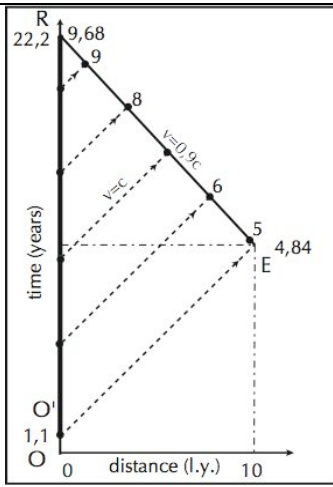 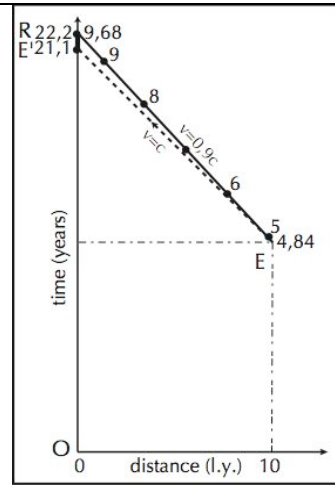

| **Fig. 3 Return Journey: What Traveller measures.** Traveller returns to Earth at R 4.84 years after arriving at E. But during this time, he observes 21.1 years elapse on Earth. **Conclusion: Traveller sees Homebody's clock beating 4.36 times faster.** | **Fig 4 Return Journey : What Homebody measures.** Homebody sees the entire return journey of Traveller take place in 1.1 years, and meets him at R 22.2 years after the initial departure. **Conclusion: Homebody sees Traveller's clock beating 4.36 times faster.** |
|---|---|

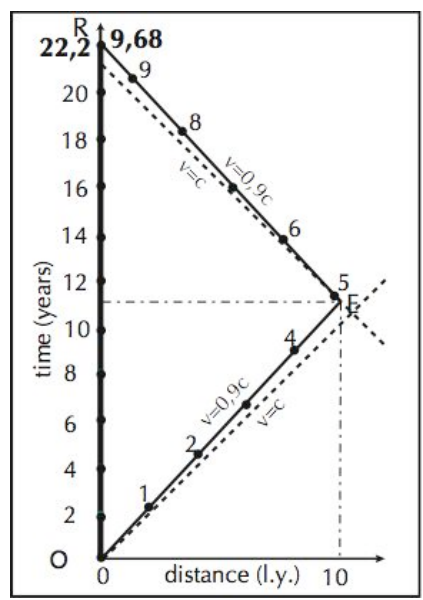

**Fig 5 : Complete Journey.** When Homebody and Traveller meet each other at R, Homebody's clock has measured 22.2 years and Traveller's clock has measured 9.68 years.

Both aspects of the paradox are solved in an obvious way by these space-time diagrams.

1) Why is the global situation not symmetric?

During the outward journey, the situations are perfectly symmetric because the inertial frames of both Traveller and Homebody are in uniform motion with relative speed *v* (fig. 1 and 2). Also, during the return journey, the situations are perfectly symmetric because the inertial frames of both Traveller and Homebody are in uniform motion with relative speed *-v* (fig. 3 and 4). But if one considers the complete journey (fig. 5), the trajectories are physically asymmetric because at E, Traveller – having modified his speed, i.e. having undergone an acceleration – changes his inertial frame.

2) Why is Traveller's proper time shorter than that of Homebody?

One can consider that it is because of the accelerations and the decelerations that Traveller has to undergo to leave Homebody at O, turn back at E and rejoin Homebody at R. Let us note however that the phases of acceleration at O and deceleration at R can be suppressed if one assumes that the trajectories of Traveller and Homebody cross without either observer stopping, their clocks being compared during the crossing. Nevertheless the necessary change of direction at E remains, translated as the acceleration of Traveller.

According to a more geometrical point of view, it is the particular structure of the relativistic space-time that is responsible for the difference of proper times. Let us see why. In classical mechanics and ordinary space, the Pythagorean theorem indicates that $\Delta Z^2 = \Delta X^2 + \Delta Y^2$, as in any right-angled triangle, which implies that $\Delta Z < \Delta X + \Delta Y$. But Special Relativity requires the introduction of a four-dimensional geometrical structure, the *Poincaré-Minkowski space-time*, which couples space and time through the speed of light (fig. 6). The Pythagorean theorem becomes $\Delta S^2 = \Delta X^2 - c^2 \Delta T^2$, and a straightforward algebraic manipulation allows us to deduce that $\Delta S$ is always longer than ΔX + cΔT. As said previously, ΔS measures the proper time. *In Poincaré-Minkowski geometry, the worldlines of inertially moving bodies maximize the proper time elapsed between two events.* One can also see that the proper time vanishes for ΔX = cΔT, in other words for *v* = ΔX / ΔT = *c*. As already pointed out, a photon never ages.

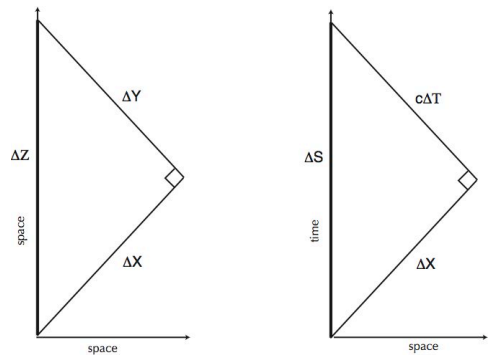

**Fig. 6 : Euclidean space (left) vs. Poincaré-Minkowski space-time (right)**

**The twin paradox in General Relativity**

General relativity deals with more realistic situations, including progressive accelerations, gravitational fields and curved space-time. The inertial frames are now systems in free-fall, and the equations which allow us to pass between inertial systems are no longer the Lorentz transformations, but the Poincaré transformations. The complete treatment of the problem of the twins within this new framework was first described by Einstein (1918), see also Perrin (1970). As in Special Relativity, the situation is never symmetric. In order to achieve his journey, Traveller necessarily experiences a finite and varying acceleration; thus he switches from one inertial reference frame to another, and his state of motion is not equivalent to that of Homebody. The rule stays the same: the twin who travelled through several inertial frames will always have aged less than the twin who stayed in the same inertial frame.

Assume for instance that the spaceship has a constant acceleration with respect to its instantaneous inertial reference frame, equal to the acceleration due to gravity at the Earth's surface and thus quite comfortable for Traveller. The spaceship velocity will rapidly increase and approach the speed of light without ever reaching it. On board, time will pass much more slowly than on Earth. In 2.5 years as measured by his own clock, Traveller will reach the closest star (Alpha Centauri) which is 4 light years from Earth, and after about 4.5 years he will have travelled 40 light-years, while his homebody twin will have died of old age. The centre of the Galaxy will be reached in 10 years, but 15,000 years would have passed on Earth. In about 30 years of his proper lifetime, the Traveller would be able to cross once around the observable Universe, that is one hundred thousand million light years! It would be better therefore not to return to Earth, since the Sun would have been extinguished long ago, after having burnt the planets to a cinder…

This shows in passing that, contrary to popular belief, although the theory of relativity prevents us from travelling faster than the velocity of light, it does facilitate the exploration of deep space. This fantastic journey is, however, impossible to realise because of the enormous amount of energy required to maintain the spaceship's acceleration. The best method would be to transform the material of the ship itself into propulsive energy. With perfectly efficient conversion, upon arrival at the centre of the Galaxy only one billionth of the initial mass would remain. A mountain would have shrunk to the size of a mouse!

The full general relativistic calculations, although less straightforward than in Special Relativity, do not pose any particular difficulty, but must take into account the fact that time acquires an additional elasticity: gravity also slows down clocks. Thus there is an additional *gravitational time dilation*, given by (1 + $\Phi/c^2$), where $\Phi$ is the difference in gravitational potentials. For instance, a clock at rest on the first floor beats more slowly than a clock at rest on the second floor (although the difference is tiny).

Physicists have been able to design clocks precise enough to experimentally test the twin paradox in a gravitational field as weak as that of the Earth. In 1971, the US Naval Observatory placed extremely precise cesium clocks aboard two planes, one flying westward and the other eastward. Upon their return, the flying clocks were compared with a twin (i.e. initially synchronized) clock kept at rest in a lab on Earth. In this experiment, two effects entered the game: a Special Relativistic effect due to the speed of the planes (about 1000 kph), and a General Relativitivistic effect due to the weaker gravity on board the planes. The clock which had travelled westward advanced by 273 billionths of a second, the one that had travelled eastward delayed by 59 billionths of a second - in perfect agreement with the fully relativistic calculations (see Hafele and Keating, 1972).

Nevertheless, it is pointless to dream of extending one's lifetime by traveling. If a human being spent 60 years of his life on board a plane flying at a velocity of 1000 kph, he would gain only 0.001 second over those who remained on the ground... (and would probably lose several years of his life due to stress and sedentarity!)

**Play Space**

Is acceleration, which introduces an asymmetry between the reference frames, the only explanation of the twin paradox? The answer is no, as many authors have pointed out. See e.g. Peters (1983), who considers the example of non–accelerated twins in a space with a compact dimension (the closure being due to non zero curvature or to topology). In such a case, the twins can meet again with neither of them being accelerated, yet they age differently!

Before revisiting the question in such a framework, let us get some insight on the global properties of space. Topology is an extension of geometry that deals with the nature of space, investigating its overall features, such as its number of dimensions, finiteness or infiniteness, connectivity properties or orientability, without introducing any measurement.

Of particular importance in topology are functions that are continuous. They can be thought of as those that stretch space without tearing it apart or gluing distinct parts together. The topological properties are just those that remain insensitive to such deformations. With the condition of not cutting, piercing, or gluing space, one can stretch it, crush it, or knead it in any way, and one will not change its topology, e.g., whether it is finite or infinite, whether it has holes or not, the number of holes if it has them, and so on. For instance, it is easy to see that, although continuous deformations may move the holes in a surface, they can neither create nor destroy them.

All spaces which can be continuously deformed one into another have the same topology. For a topologist, a ring and a coffee cup are one and the same object, characterized by a hole through which one can pass one's finger (although it is better not to pour coffee into a ring). On the other hand, a mug and a bowl, which may both serve for drinking, are radically different on the level of topology, since a bowl does not have a handle.

To speak only about 3-dimensional spaces of Euclidean type (with zero curvature), there are eighteen different topologies. Apart from the usual, infinite Euclidean space, the 17 others can be obtained by identifying various parts of ordinary space in different ways.

To make the description easier, it is useful to consider the more visualisable context of 2 dimensional spaces (i.e. surfaces). Besides the usual infinite Euclidean plane, there are four other Euclidean surfaces (figure 7). The cylinder is obtained by gluing together the opposite sides of an infinite strip with parallel edges, and Möbius band by twisting an edge through 180° before gluing the edges in the same way. The torus is obtained by gluing the opposite edges of a rectangle, and Klein's bottle by twisting one pair of edges before gluing.

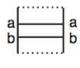

**Fig. 7. The four multiply connected topologies of the 2-dimensional Euclidean plane**. They are constructed from a rectangle or an infinite band (the fundamental domain) by identification of opposite edges according to allowable transformations. We indicate their overall shape, compactness and orientability properties.

All these surfaces have no intrinsic curvature - the sum of the angles of a triangle is always equal to 180°. They are only bent in a third dimension, which cannot be perceived by a 2-dimensional being living on the plane. Such surfaces are said to be locally Euclidean.

The rectangle we start with is called the "fundamental domain". The geometrical transformations which identify the edge-to-edge points define the way objects move continuously within this space, leaving the rectangle by an edge immediately to reappear at the other edge.

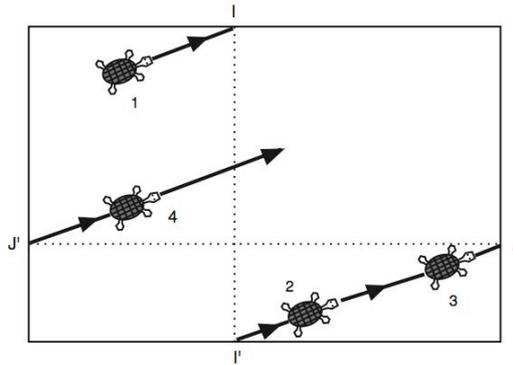

**Fig. 8: Walking on a torus**

As in those video games where characters who leave at an edge return at the opposite edge, the tortoise crosses the top of the square at *I*, reappears at the bottom at the equivalent point *I* ', continues to travel in a straight line, reaches the right-hand edge at *J*, reappears at *J* ', and so on. The torus is thus equivalent to a rectangle with the opposite edges "glued together".

One can visualize the metric properties of the space by duplicating the fundamental domain a large number of times. This generates the *universal covering space*, in which every point is repeated as often as the domain itself. One can draw in the universal covering space the various paths connecting a point to itself, called loops, either by going out of the fundamental domain to join a duplicate, in which case it is a loop which "goes around" space, or by returning towards the original point in the fundamental domain, in which case it is a loop which can be continuously shrunk to a point (figure 9).

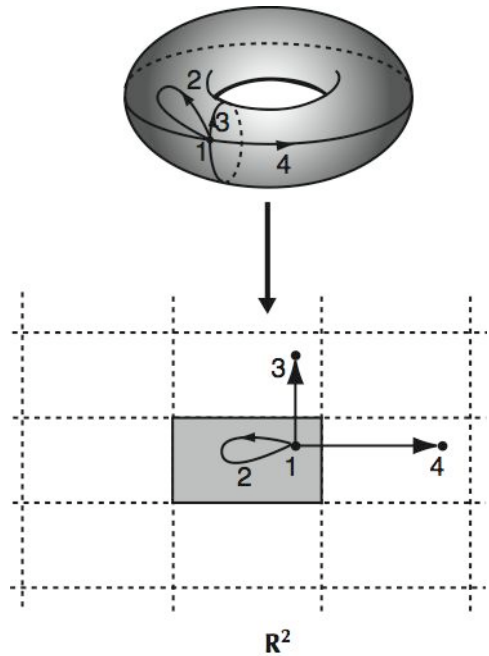

**Fig. 9: From multiply connected space to the universal covering space**

The fundamental domain of the torus is a rectangle. By repeatedly duplicating the rectangle, one generates the universal covering space - here the Euclidean plane $R^2$. The paths 2, 3 and 4 all connect the point 1 to itself. Loop 2 can be shrunk to a point, loops 3 and 4 cannot because they go around the space.

**The twin paradox in finite space**

Now we can revisit the twin paradox whatever the global shape of space may be (Barrow and Levin, 2001; Uzan et al., 2002). The travelling twin can remain in an inertial frame for all time as he travels around a compact dimension of space, never stopping or turning. Since both twins are inertial, both should see the other suffer a time dilation. The paradox again arises that both will believe the other to be younger when the twin in the rocket flies by. However, the calculations show that the twin in the rocket is younger than his sibling after a complete transit around the compact dimension.

The resolution hinges on the existence of a new kind of asymmetry between the spacetime paths joining two events, an asymmetry which is not due to acceleration but to the multiply connected topology. As we explain below, all the inertial frames are not equivalent, and the topology introduces a preferred class of inertial frames.

For the sake of visualization, let us develop our reasoning in a two-dimensional Euclidean space only (plus the dimension of time), and select the case of the flat torus. Our conclusions will remain valid in 4-dimensional space-times, whatever the topology and the (constant) spatial curvature may be.

To span all possible scenarios, let us widen the example of the twins to a family of quadruplets (strictly speaking, initially synchronized clocks) labeled 1, 2, 3 and 4 (see figure 10). 1 stays at home, at point O, and his worldline can be identified with the time axis, so that he "arrives" at O' on the space-time diagram. 2 leaves home at time t=0, travels in a rocket, turns back and joins his sibling 1 at O'. 3 and 4 also leave at time $t = 0$ but travel in different directions along non-accelerated, straight worldlines issuing from O. 3 travels along a circumference around the main line of the torus, while 4 travels around the small axis. After a while they reach points O'' and O''' respectively, and since space is closed and multiply connected, all the quadruplets meet at the same point O'. Now, one wants to compare the ages of the quadruplets when they meet.

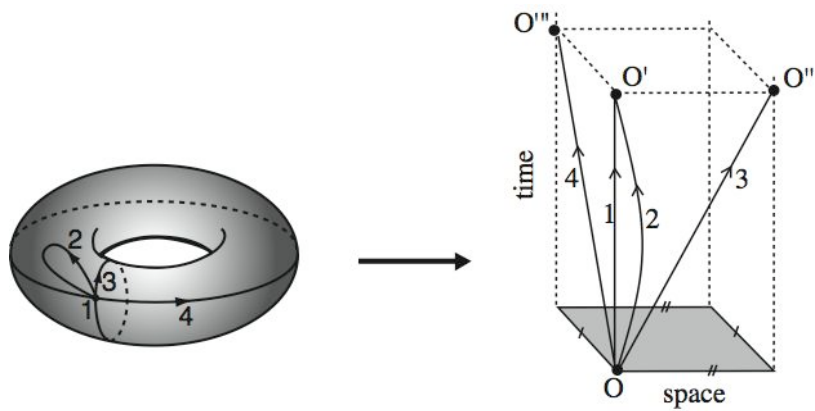

**Fig. 10: From space to space-time**

FIG. 2. On the right plot, quadruplets in a space-time with toroidal spatial sections leave O at the same time. While 1 remains at home, 2 goes away and then comes back to meet 1 at O' (corresponding to the standard case), 3 goes around the universe in a given direction from O to O'' and 4 goes around the universe along another direction from O to O'''. On the left plot, we depict the spatial projections of their trajectories on the torus. The space-time events O, O', O'', O''' are projected onto the same base point 1.

The motion of 2 corresponds to the standard paradox. Since he followed an accelerated worldline, he is younger than his sibling 1.

But there seems to be a real paradox with 1, 3 and 4, who all followed strictly inertial trajectories. Despite this, 3 and 4 are also younger than 1. In fact, the homebody 1 is always older than any traveller, because his state of motion is not symmetrical with respect to those of non–accelerated travellers.

What kind of asymmetry is to be considered here? The only explanation lies in a global breakdown of symmetry due to the multiply connected topology. Let us investigate the case more closely. If one draws closed curves (i.e. loops) on a given surface, there are two possibilities. First, the loop can be tightened and continuously reduced to a point without encountering any obstacle. This is the case for all loops in the Euclidean plane or on the sphere, for instance, and such surfaces are called simply connected. Second, the loop cannot be tightened because it goes around a "hole", as in the case of the cylinder or the torus. Such surfaces are said to have a *multiply connected* topology (multiple connectivity appears as soon as one performs gluings, or identifications of points, in a simply connected space).

Two loops are *homotopic* if they can be continuously deformed into one another. Homotopy allows us to define classes of topologically equivalent loops. In our example, the trajectories of brothers 1 and 2 are homotopic to {0}, because they can both be deformed to a point. However, they are not symmetrical because only 1 stays in an inertial frame. Here the asymmetry is due to acceleration. One can show that among all the homotopic curves from O to O', only one corresponds to an inertial observer, and it is he who will age most, as expected in the standard twin paradox.

Now, 3 and 4 travel once around the hole and once around the handle of the torus respectively. From a topological point of view, their paths are not homotopic; they can be characterized by a so–called winding index. In a cylinder, the winding index is just an integer which counts the number of times a loop goes around the surface. In the case of a torus, the winding index is a couple (m, n) of integers where m and n respectively count the numbers of times the loop goes around the hole and the handle. In our example, 1 and 2 have the same winding index (0, 0), whereas 3 and 4 have winding indices of (1, 0) and (0, 1) respectively. The winding index is a topological invariant for each traveller: neither change of coordinates nor of reference frame can change its value.

To summarize, we have the following two situations:

1. Two brothers have the same winding index (1 and 2 in our example), because their loops belong to the same homotopy class. Nevertheless only one (twin 1) can travel from the first meeting point to the second without changing inertial frame. The situations relative to 1 and 2 are not symmetrical due to local acceleration, and 1 is older than 2. Quite generally, between two twins of same homotopy class, the oldest one will always be the one who does not undergo any acceleration.

2. Several brothers (1, 3 and 4 in our example) can travel from the first meeting point to the second one at constant speed, but travel along paths with different winding indices. Their situations are not symmetrical because their loops belong to different homotopy classes: 1 is older than both 3 and 4 because his path has a zero winding index.

For observers to have the same proper times it is not sufficient that their movements are equivalent in terms of acceleration, their worldlines should also be equivalent in terms of homotopy class. Among all the inertial travellers, the oldest sibling will always be the one whose trajectory is of homotopy class {0}. The spatial topology thus imposes privileged frames among the class of all inertial frames, and even if the principle of relativity remains valid locally, it is no longer valid at the global scale. This is a sign that the theory of relativity is not a global theory of space-time.

This generalises previous works, e.g. Brans and Stewart (1973), Low (1990), Dray (1990), by adding topological considerations that hold no matter what the shape of space is.

**The complete solution**

In order to exhaustively solve the twin paradox in a multiply connected space, one would like not only to separately compare the ages of the travellers with the age of the homebody, but also to compare the ages of the various travellers when they meet each other. It is clear that only having knowledge of the winding index or the homotopy class of their loops does not allow us, in general, to compare their various proper time lapses. The only exception is that of the cylinder, where a larger winding number always corresponds to a shorter proper time lapse. But for a torus of unequal lengths, for instance when the diameter of the hole is much larger than the diameter of the handle, a traveller may go around the handle many times with a winding index (0, n), and yet be older than the traveller who goes around the hole only once with a winding index (1, 0). The situation is still more striking with a double torus, a hyperbolic surface rather than a Euclidean one (see e.g. Lachièze-Rey & Luminet, 1995). The winding indices become quadruples of integers and, as in the case of the simple torus, they cannot be compared to answer the question about the ages of the travellers. As we shall now see, this problem can only be solved by acquiring additional metric information.

The torus is built from a rectangle by gluing together its opposite sides. If one repeatedly duplicates this rectangle so as to cover the plane, one generates the universal covering space, which is infinite in all directions. It is a fictitious space that represents space as it appears to an observer located at O. All points O are, however, identical. In this representation, the trajectory of 2 appears as a loop which returns to the inital point O, without passing through one of its duplicates, whereas the trajectories of 3 and 4 are straight lines which connect the point O to a duplicate with winding numbers (1, 0) and (0, 1) respectively. There are many ways to describe a loop in a closed space, and one could consider trajectories 5 and 6 with winding numbers (1, 1) and (2, 1), for example.

As mentioned above, the homotopy classes only tell us which twin is aging the fastest: the one who follows a straight loop homotopic to {0}. They do not provide a ranking of the ages (i.e. proper time lengths) along all straight loops. To do this, some additional information is necessary, such as the various identification lengths. Indeed there exists a simple criterion which works in all cases: a shorter spatial length in the universal covering space will always correspond to a longer proper time. To fully solve the question, it is therefore sufficient to draw the universal covering space as tessellated by duplicates of the fundamental domain, and to measure the lengths of the various straight paths joining the position of sibling 1 in the fundamental domain to his ghost positions in the adjacent domains (see figure 11). As usual in topology, all reasoning involving metrical measurements can be solved in the simply–connected universal covering space.

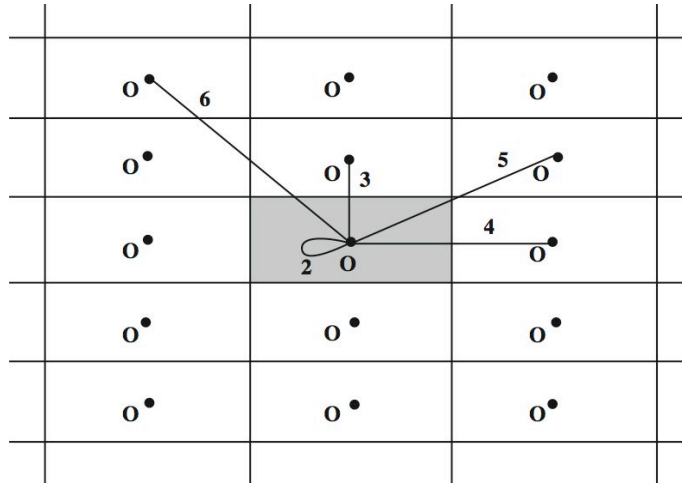

Fig. 11. Straight paths in the universal covering space of a (2 + 1)–spacetime with flat, torus–like spatial sections. Path 2 is an accelerated, curved loop with winding index (0, 0). Paths 3, 4, 5 and 6 are straight loops with respective winding indices (0, 1), (1, 0), (1, 1) and (1, 2), allowing the travellers to leave and return to the homebody at O without accelerating. The inertial worldlines are clearly not equivalent: the longer the spatial length in the universal covering space, the shorter the proper time traversed in space-time.

**The twin paradox and broken symmetry groups**

With the homotopy class, we have found a topological invariant attached to each twin's worldline which accounts for the asymmetry between their various inertial reference frames. Why is this? In Special Relativity theory, two reference frames are equivalent if there is a Lorentz transformation from one to the other. The set of all Lorentz transformations is called the Poincaré group – a ten dimensional group which combines translations and homogeneous Lorentz transformations called "boosts". The loss of equivalence between inertial frames is due to the fact that a multiply connected spatial topology globally breaks the Poincaré group.

The preceding reasoning involved Euclidean spatial sections of space-time.

In the framework of General Relativity, general solutions of Einstein's field equations are curved spacetimes admitting no particular symmetry. However, all known exact solutions admit symmetry groups (although less rich than the Poincaré group). For instance, the usual "big bang" cosmological models – described by the Friedmann–Lemaître solutions – are assumed to be globally homogeneous and isotropic. From a geometrical point of view, this means that spacelike slices have constant curvature and that space is spherically symmetric about each point. In the language of group theory, the space-time is invariant under a six-dimensional isometry group. The universal covering spaces of constant curvature are either the usual Euclidean space $R^3$, the hypersphere $S^3$, or the hyperbolic space $H^3$, depending on whether the curvature is zero, positive or negative. Any identification of points in these simply-connected spaces via a group of continuous transformations lowers the dimension of their isometry group; it preserves the three–dimensional homogeneity group (spacelike slices still have constant curvature), but it globally breaks the isotropy group (at a given point there are a discrete set of preferred directions along which the universe does not look the same).

Thus in Friedmann–Lemaître universes, (i) the expansion of the universe and (ii) the existence of a multiply connected topology for the constant time hypersurfaces both break the Poincaré invariance and single out the same "privileged" inertial observer who will age more quickly than any other twin – the one comoving with the cosmic fluid – although aging more quickly than all his travelling brothers may not be a real privilege!